\documentclass[letterpaper,journal]{IEEEtran}
\pdfoutput=1

\usepackage[utf8]{inputenc}
\usepackage[T1]{fontenc}
\usepackage{float}
\usepackage{amsmath,amssymb,amsthm}
\usepackage{commath} 
\usepackage{etoolbox}
\usepackage{microtype}
\usepackage{threeparttable}
\usepackage{siunitx}
\usepackage{xfrac}
\usepackage{subfig}
\usepackage{multirow}
\usepackage{booktabs}
\usepackage[super]{nth}

\usepackage[bookmarks=false,hidelinks]{hyperref}

\usepackage{RobStd}
\usepackage{algorithm}
\usepackage{algpseudocode}

\usepackage{fixltx2e}

\graphicspath{{fig/}}

\newcommand{\figinput}[1]{%
\providecommand{\fns}{\footnotesize}%
\providecommand{\sss}{\scriptsize}%
\providecommand{\ts}{\tiny}%
{\fontsize{10}{12}\selectfont{\textsf{\input{fig/#1}}}}%
\let\fns\undefined%
\let\ss\undefined
\let\ts\undefined
}

\newcommand{\apdxref}[1]{\hyperref[#1]{Appendix~\ref{#1}}}

\robustify\bfseries
\robustify\itshape

\DeclareMathOperator{\B}{B}

\DeclareMathOperator{\Geom}{Geom}
\DeclareMathOperator{\Beta}{Beta}

\renewcommand{\texttilde}{\raise.17ex\hbox{$\scriptstyle\sim$}}

\newcommand{\appropto}{\mathrel{\vcenter{
  \offinterlineskip\halign{\hfil$##$\cr
    \propto\cr\noalign{\kern2pt}\sim\cr\noalign{\kern-2pt}}}}}

\HideNotes

\title{Performance Analysis of Location Profile Routing}

\author{ David R.\ Bild, Yue Liu, Robert P.\ Dick, Z.\ Morley Mao, and
  Dan S.\ Wallach%
  \thanks{This work was supported by NSF under award TC-0964545.}%
  \thanks{D.\ R.\ Bild, Y.\ Liu, R.\ P.\ Dick, and Z.\ Morley Mao are
    with the Electrical Engineering and Computer Science Department,
    University of Michigan, Ann Arbor, MI 48109. E-mail:
    drbild,liuyue,dickrp,zmao@umich.edu}%
  \thanks{D.\ S.\ Wallach is with the Department of Computer Science,
    Rice University, Houston, TX 77005. E-mail: dwallach@cs.rice.edu}}

\begin{document}

\maketitle

\thispagestyle{plain}
\pagestyle{plain}

\begin{abstract}
  We propose using the predictability of human motion to eliminate the
  overhead of distributed location services in human-carried MANETs,
  dubbing the technique \emph{location profile routing}. This method
  outperforms the Geographic Hashing Location Service when nodes
  change locations $2\times $more frequently than they initiate
  connections (e.g., start new TCP streams), as in applications like
  text- and instant-messaging. Prior characterizations of human
  mobility are used to show that location profile routing achieves a
  93\% delivery ratio with a 1.75$\times$ first-packet latency
  increase relative to an oracle location service.
\end{abstract}

\section{Introduction}
\label{sec:introduction}

\Note{Routing is hard is mobile networks}

Traditional routing protocols rely on shared global state and thus
scale poorly in mobile ad hoc networks (MANETs) with frequent changes
in topology. Routing overhead grows quadratically in the number of
nodes for distance vector and link state protocols~\cite{jacquet01dec}
that must distribute changes to all nodes.  The natural hierarchy used
to reduce the overhead traffic in networks like the Internet (e.g.,
CIDR) is not available. On-demand
methods~\cite{abolhasan04jan,perkins99feb,johnson96} delay routing
table updates until needed, but only reduce overhead by constant
factors---the scaling behavior is unchanged.\footnote{We assume that
  sender and receiver locations are not correlated; that could change
  the scaling behavior.} In contrast, stateless protocols that use
local information to make forwarding decisions have the potential to
scale.

\Note{GPRS ``solves'' routing by shifting complexity to addressing}

One stateless protocol, Greedy Perimeter Stateless Routing
(GPSR)~\cite{karp00aug}, uses geography: messages are addressed to
specific locations.  Nodes already know their own locations (e.g., via
GPS), allowing each intermediate step to bring the message closer to
its destination. No global routing state is needed. Essentially
though, this technique just shifts the complexity from routing to
addressing. A forwarding node only needs its own locally-known
location, but the original sender requires the current location of the
recipient, a global mapping.

\Note{Distributed location services also don't scale.}

Distributed location services~\cite{li00aug,das05mar} can maintain
this identity to location mapping, but also have drawbacks.  Hierarchy
is imposed to manage scalability, but overhead still increases
super-linearly~\cite{das05mar}.  Further, locations are sensitive
information, so complicated schemes are required to protect privacy
and anonymity~\cite{beresford03jan}.  We observe that if node
locations are predictable, the mapping can be done locally as well,
reducing the scaling and privacy concerns.

In fact, human locations are highly regular with $\sim$93\%
predictability~\cite{song10feb}.  In MANETs of human-carried devices,
predicative models of future locations can be pre-shared among trusted
participants. These models combined with GPSR allow low-overhead
addressing and routing, with network scalability limited only by the
actual traffic. We name this approach \emph{location profile routing}
(LPR)~\cite{bild11jun} and study its performance potential. We
determine the number of locations that must be addressed to achieve
the peak 93\% packet delivery rate and derive the associated latency
and traffic overheads. Finally, we determine the conditions under
which LPR outperforms the Geographic Hashing Location Service.

\section{Description of Location Profile Routing}
\label{sec:lpr}

\begin{figure*}
  \includegraphics[width=\linewidth]{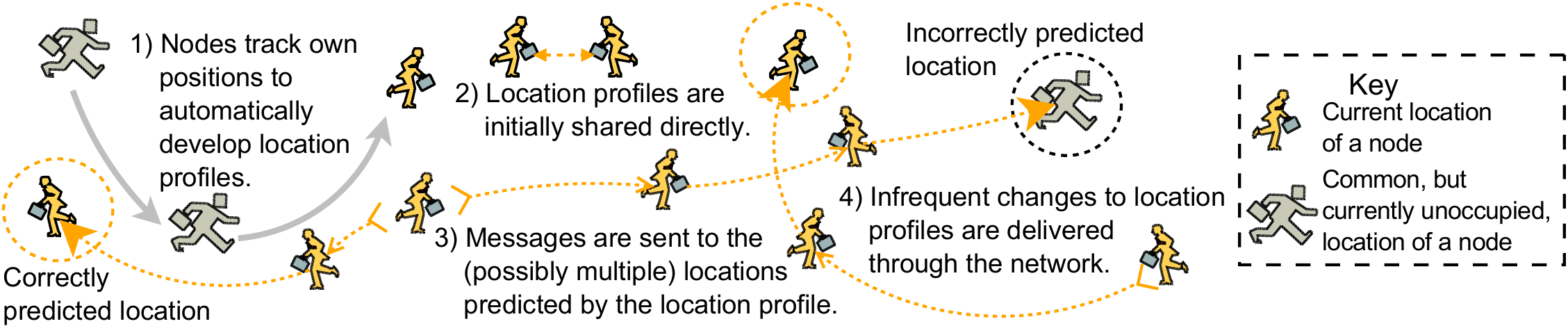}
  \caption{Illustration of the main components in location profile
    routing~\cite{bild11jun}.}
  \label{fig:location-profile-routing}
\end{figure*}

\Note{Summary of the technique, referring to figure}

Location profile routing (LPR) stems from the observation that humans
generally have simple, repeated motions, with most time spent at a few
common time-dependent locations~\cite{gonzalez08jun} easily captured
by a compact predictive model.  For the many potential\footnote{Ad hoc
  networks are not yet widely used by the general public.}
applications of human-carried MANETs that can tolerate the a reduction
in delivery reliability or increase in latency (we previously
described a particularly compelling application---censorship-resistant
personal communication~\cite{bild11jun}), LPR eliminates overhead
traffic for route maintenance.

\autoref{fig:location-profile-routing} illustrates the main steps of
LPR. Nodes continuously monitor their positions to build location
profiles (step 1), which are then shared with potential future
contacts directly (step 2).  This sharing happens a priori when two
nodes are directly connected (i.e., one hop apart), limiting bandwidth
usage.  A message is addressed to the location(s) predicted by the
profile (step 3) and delivered via GPSR. Routing fails if none of
predicted locations are correct, but delay-tolerant delivery is a
possible fallback. Changes to the motion patterns are rare (e.g., when
someone starts a new job or moves to a new home) and thus distributed
via the network (step 4).

\Note{Description of location profiles}

\textbf{Location Profiles:} Motion patterns can be modeled in many
ways, but a simple discrete model is sufficient for our purposes. A
location profile is a function $P$ mapping a time interval (e.g.,
Tuesday 15:30--15:40) to a set of location--confidence tuples,
with higher confidence indicating stronger belief in the node
occupying that location at that time:
\[
P: \mathrm{time} \mapsto \{(\mathrm{loc}_1,\mathrm{conf}_1),\ldots,(\mathrm{loc}_n,\mathrm{conf}_n)\}
\]
The precise discretization level is unimportant. Both cell-tower
granularity (\SI{3}{\kilo\meter\squared}, \SI{1}{\hour}) and WiFi AP
granularity (\SI{157}{\meter\squared}, \SI{10}{\minute}) have similar
predictabilities at 93\%~\cite{song10feb} and 92\%~\cite{burbey08sep}.

Various implementations are possible, but for completeness we
summarize the Prediction-by-Partial-Match (PPM) approach of Burbey and
Martin~\cite{burbey08sep}. PPM is a variable-order Markov model over a
sequence $S$ of observed time-interval--location pairs, $S = \{T_1 L_1
T_2 L_2 \ldots T_n L_n\}$. This defines a probability distribution
over the next location conditioned on the prior $k$ elements of
context. In our case, prior locations are not known, so our definition
of $P$ corresponds to the first-order variant ($k=1$, i.e., context is
the current time). We briefly discuss zero- (no context) and
third-order (context includes the previous location) variants.  This
scheme captures most of the predictability (90\%~\cite{burbey08sep}
vs. the 93\% reported maximum~\cite{song10feb}).

\Note{Profile Distribution}

\textbf{Profile Distribution:} Location profiles are disseminated
a-priori and out-of-band, similar to telephone numbers or email
addresses.  For our envisioned applications---communication between
friends---the profiles can be exchanged face-to-face. In other cases,
a centralized service, similar to a telephone directory, might be
needed. Regardless, the salient point is that the profiles are known
a-priori and thus can be exchanged outside of the network. Although
changes could be disseminated out-of-band as well, in-network
propagation is feasible because updates are infrequent and sent only
to select participants (e.g., friends). Opportunistically updating
when devices are in close proximity further bounds the overhead.

\Note{Addressing Policy}

\textbf{Addressing Policy:} The addressing policy translates the
location--confidence tuples output by the profile into a message
delivery strategy specifying when and where packets will be sent.
Only one of the locations can be correct, so the order and method in
which they are tried influences the network throughput and latency
trade-off.  Their spatial correlations influence the minimum cost
routing strategy (e.g., Steiner tree) to reach all locations. The
primary focus of this paper is analyzing these performance
characteristics and trade-offs.

\Note{Fallback method}

\textbf{Fallback Method:} LPR fails outright when nodes are in
unpredictable locations, i.e., at least 7\% of the
time~\cite{song10feb}.  Although this may be tolerable for many
applications in which messages can be redelivered later, it is
non-ideal.  As this is not our focus, we omit details here, but
possible strategies include delay tolerant delivery (in-network
buffering of the message at a common location until the node's return)
or rendezvous delivery (messages are sent to a rendezvous location
which the node, when not in a predictable location, apprises of
current forwarding instructions).  Such schemes allow for reliable
delivery with average overheads still drastically lower than
traditional routing approaches.

\section{Performance Analysis}
\label{sec:analysis}

We use prior empirical studies of human motion patterns to develop
analytical models suitable for studying the performance of LPR.
Barab\'{a}si \etal studied six-month location traces of 100,000
European cellphone users~\cite{gonzalez08jun,song10feb} at cell-tower
granularity, reporting a maximum predictability of 93\%. The size and
duration of the traces make this best source to date.  To confirm that
locations are as predicable at WiFi granularity, we turn to Burbey and
Martin's study~\cite{burbey08sep} of traces from 275 WiFi users
at UCSD~\cite{mcnett05apr}. They found similar predictability, 92\%,
confirming that cellular granularity is not limiting.

\subsection{How Predictable are Common Locations?}
\label{sec:first-sucess}

\begin{figure*}
\begin{minipage}{0.32\textwidth}
  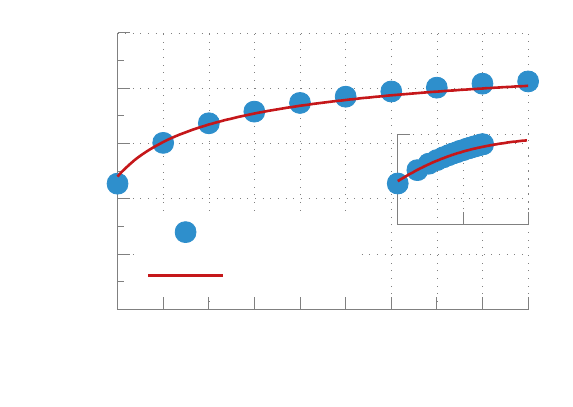
  \caption{The probability that a user currently occupies one of his
    $k$ most-common locations is well-modeled by
    \autoref{eqn:probability-success-after-n}.}
  \label{fig:probability-success-after-n}
\end{minipage}
\hfill
\begin{minipage}{0.32\textwidth}
  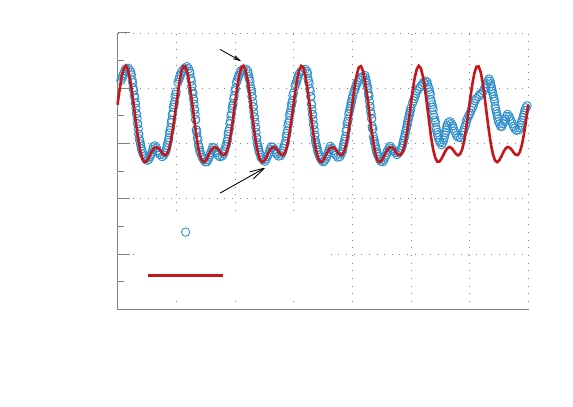
  \caption{The time-dependent regularity $R(t)$, i.e., the probability
    the user is in the most common location associated with that time
    interval.}
  \label{fig:first-location-regularity}
\end{minipage}
\hfill
\begin{minipage}{0.32\textwidth}
  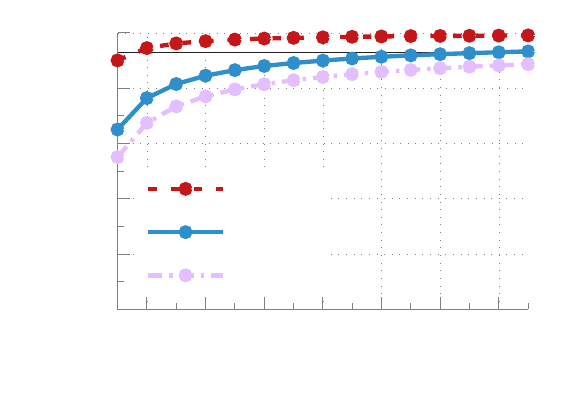
  \caption{Success rate of a first-order profile versus the number of
    locations attempted. Rates during maximum (night) and minimum
    (day) predictability are shown too.}
  \label{fig:probability-success-after-n-first-order}
\end{minipage}
\end{figure*}

A location profile returns multiple locations in order of likelihood,
so delivery cost and latency depends on how many, $K$, must be
targeted to reach the user.  Intuitively, most time is spent in two
locations---home and work---so a zero-order model (i.e., not
conditioned on current time) might be sufficient.  The pmf is
$\widetilde{\pi}(k) = p_k\prod_{i=1}^{k-1}(1-p_i),$ where $p_i$ is the
probability that the target is in location $i$. The $p_i$'s are
roughly distributed\footnote{A true Zipfian distribution requires a
  bounded domain $i\in[1,N]$ with $c = \frac{1}{H_N}$ for the $p_i$'s
  to total one. The following results are for the reported empirical
  form, not a true distribution.} as $p_i \propto i^{-1}$ with
proportionality constant $c \approx 0.48$~\cite{gonzalez08jun}.  $K$
is equivalent to a beta-geometric distribution, $K \sim \Geom(L)$ with
$L \sim \Beta(c, 1 - c)$, and has CDF
\begin{equation}
  \label{eqn:probability-success-after-n}
  \widetilde{\Pi}(k) = 1 - \frac{1}{k\B(k, 1 - c)}.
\end{equation}
The match\footnote{$L \sim \Beta(0.60, 0.72)$ yields a tighter
  fit, but lacks an explanatory origin. It \emph{might} result from
  a mixture of different upper bounds $N$ in the Zipfian model of the
  $p_i$'s---individuals have different numbers of common locations.}
to measured data~\cite{song10feb} is shown in
\autoref{fig:probability-success-after-n}. The first moment diverges,
but two locations suffice only 60\% of the time and ten achieve only
80\% delivery. Conditioning the model on time of day is necessary.

The first-order model (with \SI{10}{\minute} intervals) is 90\%
accurate for the first location on the UCSD
dataset~\cite{burbey08sep}, nearing the 93\% upper bound and
suggesting marginal gains for additional guesses. A third-order model
is surprisingly only slightly better at 92\%.  The larger cellular
dataset (with \SI{1}{\hour} intervals) is more pessimistic. The
accuracy $R(t)$ of the first-order model here is given by
\begin{equation}
\label{eqn:first-location-regularity}
R(t) = c_1\sin\left(\frac{2\pi}{24}t + \frac{2\pi}{8}\right) + 
c_2\sin\left(\frac{2\pi}{12}t - \frac{2\pi}{24}\right) + c_3,
\end{equation}
where $c_1 = 0.148$, $c_2 = 0.077$, $c_3 = 0.657$ and $t \in [0,167]$
is the hour of the week, i.e., $t = 0$ is Monday 00:00--0:59 and $t =
167$ is Sunday 23:00--23:59. As shown in
\autoref{fig:first-location-regularity}, this form captures one-day
and half-day periodicities. On weekends, the variability is lower and
the intervals of highest predictability occur later in the day The
accuracy on weekdays ranges from 55\% to 90\%, averaging $\bar{R}
\approx 65\%$.

Assuming the power law form, $p_i \propto i^{-1}$, holds during each
time interval\footnote{The number of common locations is inversely
  correlated with $R(t)$~\cite{song10feb} (Fig. 3B), suggesting that
  it does.}, equations \ref{eqn:probability-success-after-n} and
\ref{eqn:first-location-regularity} can be combined as
\begin{equation}
  \label{eqn:probability-success-after-n-first-order}
  \widetilde{\Pi}_1(k) = 1 - \int_0^{168} \frac{D(t)}{k\B(k,1-R(t))}\,dt,
\end{equation}
where $D(t)$ is the traffic density at time $t$, to yield the average
probability that packet addressed to the $k$-most common locations
reaches the target, shown in
\autoref{fig:probability-success-after-n-first-order}.  We assume a
uniform density, $D(t) = \frac{1}{168}$, but other known traffic
patterns can be substituted. $k=5$ achieves 85\% success and 93\%
requires only $k=12$. More locations are required during the day and
fewer at night.  The exact number of locations to attempt is
application-specific, depending on the trade-off between between
desired delivery rate and cost, i.e., increased latency and traffic
overhead.

\subsection{What Additional Latency and Traffic is Induced by
  LPR?}
\label{sec:additional-latency}

\begin{figure*}
\begin{minipage}{0.32\textwidth}
  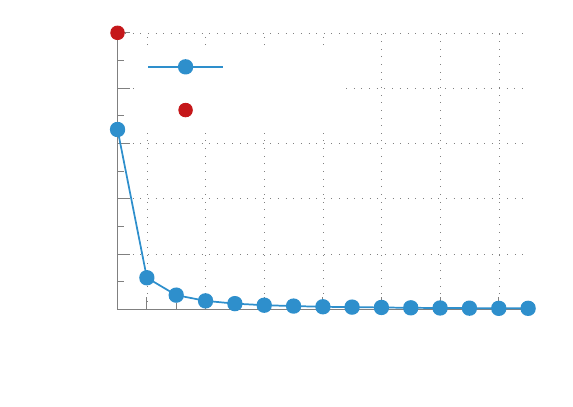
  \caption{PMF of the latency increase for the first packet in a
    stream induced by trying multiple locations in turn. Concurrent
    attempts do not impact latency.}
  \label{fig:latency-distribution}
\end{minipage}
\hfill
\begin{minipage}{0.32\textwidth}
  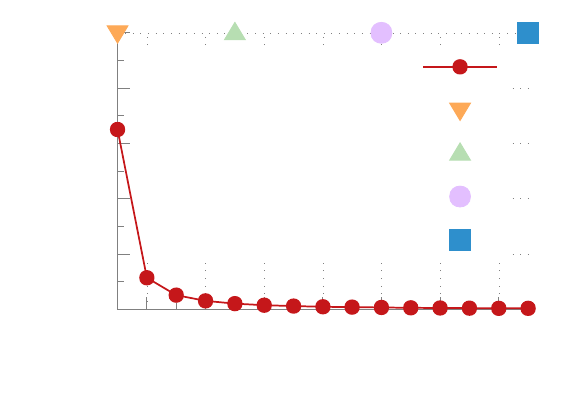
  \caption{PMF of the traffic overhead for the first packet in a
    stream induced by trying locations in turn. Concurrent attempts
    have a fixed overhead.}
  \label{fig:overhead-distribution}
\end{minipage}
\hfill
\begin{minipage}{0.32\textwidth}
  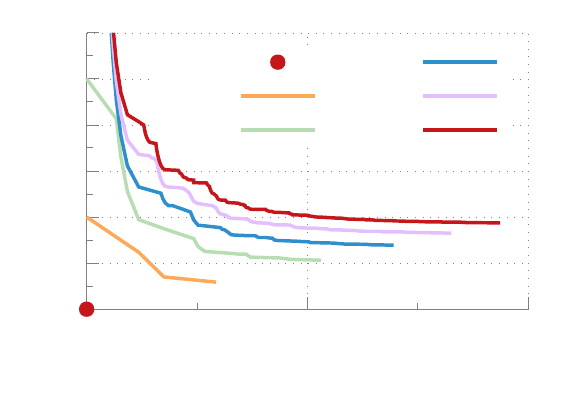
  \caption{Pareto front of the first packet latency--traffic
    trade-off of a combined parallel-series strategy for several
    average success rates.  }
  \label{fig:latency-overhead-pareto}
\end{minipage}
\end{figure*}

Some packets must be sent to multiple locations to have an adequate
packet delivery rate, increasing latency and traffic by constant
factors.  Note that the costs increase only for the first packet in a
stream. Subsequent packets are sent directly to the now-known current
location. The true average overhead depends on the percentage of first
packets, which is low for applications like text-messaging and email
and higher for interactive applications like voice chat. We report
overheads for first packets only, which readers should scale by the
first packet percentage of their applications.

We assume that receiver common locations and sender locations are
uniformly distributed in the network.\footnote{Spatially-correlated
  locations can reduce overheads (see \autoref{sec:ghls}).}  Thus, we
can report overheads relative to the average latency (round-trip time)
and traffic cost (round-trip hop count) for a single delivery attempt,
e.g., a $2\times$ increase.

Parallel delivery to all $k$ common locations does not increase
latency, but increases traffic by $k\times$.  Serial
delivery---attempting each location only if the previous failed, using
ACKs and a timeout to detect failure---reduces the traffic
overhead. The pmf of the factor increase $T$, plotted in
\autoref{fig:overhead-distribution}, is
\begin{equation}
  \Pr[T = t] = \tilde{\pi}_1(t),
\end{equation}
where $\tilde{\pi}_1$ is the pmf associated with
\autoref{eqn:probability-success-after-n-first-order}. Latency
increases similarly, as shown in \autoref{fig:latency-distribution}.

A combined approach---addressing a subset of the locations in
parallel---can fine-tune the trade-off. For example, four different
groupings can be used when trying three locations ($\sim81\%$ success
rate).

{\centering 
\begingroup%
  \makeatletter%
  \providecommand\color[2][]{%
    \errmessage{(Inkscape) Color is used for the text in Inkscape, but the package 'color.sty' is not loaded}%
    \renewcommand\color[2][]{}%
  }%
  \providecommand\transparent[1]{%
    \errmessage{(Inkscape) Transparency is used (non-zero) for the text in Inkscape, but the package 'transparent.sty' is not loaded}%
    \renewcommand\transparent[1]{}%
  }%
  \providecommand\rotatebox[2]{#2}%
  \ifx\svgwidth\undefined%
    \setlength{\unitlength}{248.4bp}%
    \ifx\svgscale\undefined%
      \relax%
    \else%
      \setlength{\unitlength}{\unitlength * \real{\svgscale}}%
    \fi%
  \else%
    \setlength{\unitlength}{\svgwidth}%
  \fi%
  \global\let\svgwidth\undefined%
  \global\let\svgscale\undefined%
  \makeatother%
  \begin{picture}(1,0.06956522)%
    \put(0,0){\includegraphics[width=\unitlength]{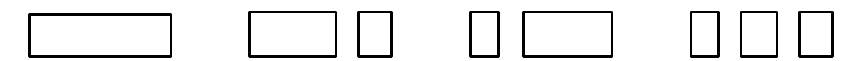}}%
    \put(0.05002412,0.01389304){\color[rgb]{0,0,0}\makebox(0,0)[b]{\smash{1}}}%
    \put(0.11443636,0.01389304){\color[rgb]{0,0,0}\makebox(0,0)[b]{\smash{2}}}%
    \put(0.17884859,0.01389304){\color[rgb]{0,0,0}\makebox(0,0)[b]{\smash{3}}}%
    \put(0.30575948,0.01389284){\color[rgb]{0,0,0}\makebox(0,0)[b]{\smash{1}}}%
    \put(0.37017172,0.01389284){\color[rgb]{0,0,0}\makebox(0,0)[b]{\smash{2}}}%
    \put(0.43458396,0.01389284){\color[rgb]{0,0,0}\makebox(0,0)[b]{\smash{3}}}%
    \put(0.56149489,0.01389303){\color[rgb]{0,0,0}\makebox(0,0)[b]{\smash{1}}}%
    \put(0.62590712,0.01389303){\color[rgb]{0,0,0}\makebox(0,0)[b]{\smash{2}}}%
    \put(0.69031934,0.01389303){\color[rgb]{0,0,0}\makebox(0,0)[b]{\smash{3}}}%
    \put(0.81723027,0.01389284){\color[rgb]{0,0,0}\makebox(0,0)[b]{\smash{1}}}%
    \put(0.88164251,0.01389284){\color[rgb]{0,0,0}\makebox(0,0)[b]{\smash{2}}}%
    \put(0.94605473,0.01389284){\color[rgb]{0,0,0}\makebox(0,0)[b]{\smash{3}}}%
  \end{picture}%
\endgroup%
\\
}
\noindent All locations within a group (a box in the diagram) are
tried concurrently and groups are tried serially from left to right,
as needed. Formally, a grouping $G$ is a partition of the common
locations, $G = \{g_1,g_2,\ldots\}$, with the property that for $i <
j$, all locations in group $g_i$ are more probable than those in
$g_j$. Let $\kappa(g)$ denote the index of the most common location in
$g$, e.g., $\kappa(g_1) = 1$. Then, the probability that group $g$ is
tried, i.e., that all previous groups failed, is $\Phi_1(g) = 1 -
\tilde{\Pi}_1(\kappa(g) - 1)$.  Thus, the average latency increase
$\bar{L}$ for a grouping $G$ is
\begin{equation}
  \bar{L}(G) = \sum_{g\in G}\Phi_1(g)
\end{equation}
and the average traffic overhead $\bar{T}$ is
\begin{equation}
  \bar{T}(G) = \sum_{g\in G}|g|\Phi_1(g).
\end{equation}
\autoref{fig:latency-overhead-pareto} shows the Pareto fronts for
several average success rates, i.e., the maximum number of locations
attempted.  At the knees, $\bar{L} \approx 1.25\times$ and $\bar{T}
\approx 3\textrm{--}4\times$. These curves are network averages. At
runtime when a specific location profile is known, the precise
trade-offs for that instance can be computed.

\subsection{Under What Conditions Does LPR Outperform Location
  Services?}
\label{sec:ghls}

LPR trades the cost of updating a location service as devices move for
multiple transmissions at the first packet. We use a simple analytical
model to derive the network conditions under which LPR outperforms the
Geographic Hashing Location Service (GHLS)~\cite{das05mar}, a scalable
distributed location service. Let $f$ be the network-wide location
update rate (which increases with node movement), $r$ be the
network-wide first-packet rate, $s$ be the average number of hops
between a node and its GHLS location server, $p$ be the average number
of hops between a source and destination, and $\bar{T}$, as previously
defined, be the average number of destinations attempted by LPR. The
location update, location query, and first-packet delivery costs
(i.e., transmission counts) for GHLS are\footnote{See Das \etal
  (Section IV)~\cite{das05mar} for the derivation. Our $s$ is their
  $\frac{1}{3}2^{h-1}\sqrt{2}$.} $fs$, $2rs$, and $2rp$. LPR has only
the first-packet delivery cost, $2\bar{T}rp$.  After rearranging the
total costs in terms of $\frac{f}{r}$ and $\frac{p}{s}$, we see that
LPR has lower overhead when
\begin{equation}
  \frac{f}{r} > \frac{p}{s}(2\bar{T}-2) - 2.
\end{equation}
When $s=p$ (source and destination are uniformly distributed over the
entire field) and $\bar{T} \approx 3$ (from
\autoref{fig:latency-overhead-pareto}), this simplifies to
$\frac{f}{r} > 2$; LPR outperforms GHLS when the location update rate
is more than twice the first-packet rate. This bound further decreases
when sources and destinations are spatially concentrated, i.e., $p <
s$.

\subsection{Reducing Overhead Via Spanning Trees}
\label{sec:steiner}

The preceding overhead and latency analysis assumed linear routing,
i.e., one transmission from the source per attempted destination. A
branching route (e.g., the Euclidean Steiner tree containing the
source and destinations) would reduce this overhead, particularly when
destinations are spatially-clustered relative to the
source. Unfortunately, this works only for dense networks in which
nodes are guaranteed to exist at the branching (e.g., Steiner)
points. Many real-world networks are too irregular, and the linear
approach should be used.

In dense networks, the branching approach is feasible. One desires a
routing tree with low total weight to minimize traffic but also with
short source-to-destination path lengths to minimize latency.
Although seemingly conflicting, both goals are achievable. Taking $n$
as the size of the network, trees with weights within $o(n)$ of the
$O(n)$-length minimal Steiner tree and source-to-destination path
lengths within $o(\log n)$ of the $O(\sqrt{n})$ straight-line
distances exist~\cite{aldous08mar}. We refer the reader to Aldous and
Kendall for details and construction~\cite{aldous08mar}.

\section{Conclusion}
\label{sec:conclusion}

We have argued that predicable motion patterns can replace expensive
distributed location services in human-carried MANETs, reducing
overhead transmissions.  The promising potential of LPR highlights the
advantages of considering deployment- and application-specific
behaviors in lower levels of the network stack. Our analysis focused
on MANETs, but the approach applies equally to delay-tolerant
networks. In light of growing commercial interest in device-to-device
communication for smartphones (e.g., Wi-Fi Direct), we hope this work
spurs further interest in adopting human behavior to improve ad hoc
network performance.

\bibliographystyle{IEEEtran}
\bibliography{robbib,rdgroup}

\end{document}